\documentclass[pra,twocolumn,nofootinbib,floatfix,10pt]{revtex4-2}

\usepackage{amsmath}
\usepackage{amssymb}
\usepackage{wasysym}
\usepackage{graphicx}
\usepackage{color,soul}
\usepackage{physics}
\usepackage{siunitx}
\usepackage{dsfont}
\usepackage{float}

\usepackage[english]{babel}
\usepackage{blindtext}
\usepackage[hidelinks]{hyperref}

\usepackage[english,nomargin,inline,marginclue,draft]{fixme}
\fxusetheme{colorsig}
\FXRegisterAuthor{cg}{acg}{CG}  
\FXRegisterAuthor{th}{ath}{\color{blue}TH}  
\FXRegisterAuthor{ib}{aib}{\color{red}IB} 
\FXRegisterAuthor{sh}{ash}{\color{cyan}SH} 
\FXRegisterAuthor{db}{adb}{\color{green}DB} 
\FXRegisterAuthor{ps}{aps}{PS} 
\makeatletter
\renewcommand*\FXLayoutInline[3]{%
  {\@fxuseface{inline}\ignorespaces{\color{fx#1}[#3: #2]}}}
\makeatother

\long\def\symbolfootnote[#1]#2{\begingroup%
\def\thefootnote{\fnsymbol{footnote}}\footnotetext[#1]{#2}\endgroup}

\def\nobreakbefore{%
  \relax\ifvmode\else
    \ifhmode
      \ifdim\lastskip > 0pt\relax
        \unskip\nobreakspace
      \else 
        \nobreakspace
      \fi
    \fi
  \fi
}
\let\oldcite\cite
\renewcommand\cite{\nobreakbefore\oldcite}

\newcommand{\Jpl}{\ensuremath{J_{\parallel}}} 
\newcommand{\Jp}{\ensuremath{J_{\perp}}} 
\newcommand{\gI}{\ensuremath{g_{\mathds{1}, R^z}}}
\newcommand{\gS}{\ensuremath{g_{S^z, R^z}}} 
\newcommand{\gtS}{\ensuremath{\tilde{g}_{S^z, R^z}}} 
\newcommand{\kB}{\ensuremath{k_{\rm B}}} 
\newcommand{\gSzP}{\ensuremath{g_{S^z,P^\downarrow}}}
\newcommand{\gIP}{\ensuremath{g_{\mathds{1},P^\downarrow}}}
\newcommand{\gtP}{\ensuremath{\tilde{g}_{S^z,P^\downarrow}}}

\begin{document}

\title{Realising the Symmetry-Protected Haldane Phase in Fermi-Hubbard Ladders}

\author{Pimonpan~Sompet$^{1,2,3,\ast,\dag}$}
\author{Sarah~Hirthe$^{1,2,\ast}$}
\author{Dominik~Bourgund$^{1,2,\ast}$}
\author{Thomas~Chalopin$^{1,2}$}
\author{Julian~Bibo$^{2,4}$}
\author{Joannis~Koepsell$^{1,2}$}
\author{Petar~Bojovi\'{c}$^{1,2}$}
\author{Ruben~Verresen$^{7}$}
\author{Frank~Pollmann$^{2,4}$}
\author{Guillaume~Salomon$^{1,2,5,6}$}
\author{Christian~Gross$^{1,2,8}$}
\author{Timon~A.~Hilker$^{1,2}$}
\author{Immanuel~Bloch$^{1,2,9}$}

\affiliation{$^1$Max-Planck-Institut f\"{u}r Quantenoptik, 85748 Garching, Germany}
\affiliation{$^2$Munich Center for Quantum Science and Technology, 80799 Munich, Germany}
\affiliation{$^3$Research Center for Quantum Technology, Faculty of Science, Chiang Mai University, Chiang Mai, 50200, Thailand}
\affiliation{$^4$Department of Physics, Technical University of Munich, 85748 Garching, Germany}
\affiliation{$^5$Institut f\"{u}r Laserphysik, Universit\"{a}t Hamburg, 22761 Hamburg, Germany}
\affiliation{$^6$The Hamburg Centre for Ultrafast Imaging, Universit\"{a}t Hamburg, 22761 Hamburg, Germany}
\affiliation{$^7$Department of Physics, Harvard University, Cambridge, MA 02138, USA}
\affiliation{$^8$Physikalisches Institut, Eberhard Karls Universit\"{a}t T\"{u}bingen, 72076 T\"{u}bingen, Germany}
\affiliation{$^9$Fakult\"{a}t f\"{u}r Physik, Ludwig-Maximilians-Universit\"{a}t, 80799 M\"{u}nchen, Germany}

\symbolfootnote[1]{These authors contributed equally to this work.}
\symbolfootnote[2]{Electronic address: {\bf pimonpan.sompet@mpq.mpg.de}}


\begin{abstract}
	Topology in quantum many-body systems has profoundly changed our understanding of quantum phases of matter. The paradigmatic model that has played an instrumental role in elucidating these effects is the antiferromagnetic spin-1 Haldane chain \cite{Haldane:1983,haldane:2018}.
	Its ground state is a disordered state, with symmetry-protected fourfold-degenerate edge states due to fractional spin excitations. In the bulk, it is characterised by vanishing two-point spin correlations, gapped excitations, and a characteristic non-local order parameter\cite{dennijs:1989,kennedy:1992}.
	More recently it was understood that the Haldane chain forms a specific example of a more general classification scheme of symmetry protected topological (SPT) phases of matter that is based on ideas connecting to quantum information and entanglement \cite{schuch:2011a,pollmann:2010,chen:2011c}.
	Here, we realise such a topological Haldane phase with Fermi-Hubbard ladders in an ultracold-atom quantum simulator.
	We directly reveal both edge and bulk properties of the system through the use of single-site and particle-resolved measurements as well as non-local correlation functions. Continuously changing the Hubbard interaction strength of the system allows us to investigate the robustness of the phase to charge (density) fluctuations far from the regime of the Heisenberg model employing a novel correlator.	
\end{abstract}
\maketitle

Topological phases of matter often share a deep connection between their bulk and edge properties \cite{wen:2017,senthil:2015}.
In the case of the Haldane chain, the bulk exhibits a hidden antiferromagnetic (AFM) order characterised by AFM correlations interlaced with an arbitrary number of $S^z=0$ elements. This pattern can only be revealed through non-local string correlations that are sensitive to the local spin states, requiring however a detection of the quantum many-body system with microscopic resolution.
Even though this was not possible in early experiments on spin-1 chains, evidence for a spin-gap as well as spin-1/2 localised edge states was found using neutron scattering \cite{renard:1987,buyers:1986} or electron resonance experiments \cite{hagiwara:1990, glarum:1991} while not directly probing this hidden order or spatially resolving the edge states.
Recent developments in quantum simulations allow one to go beyond such solid-state bulk measurements by observing quantum many-body systems with single-site resolution \cite{bakr:2009,sherson:2010,haller:2015,cheuk:2015,parsons:2015} and in a fully spin- and density-resolved way\cite{boll:2016,koepsell:2020}. 
This provides a rich diagnostic tool to obtain a direct microscopic picture of the hidden order in experiments\cite{endres:2011,hilker:2017}.
The power of this technique has also been demonstrated recently in a study that was able to reveal a SPT phase in the hardcore boson Su-Schrieffer-Heeger (SSH) model using Rydberg atoms\cite{leseleuc:2019}. Here we expand on those results by realising the Haldane phase in a spin system with tunable coupling strength, system size and controlled charge fluctuations. We show this by measuring both topological and trivial string order parameters.

An instructive way to engineer the Haldane phase in systems of spin-1/2 fermions is based on the celebrated AKLT model \cite{affleck:1987,kennedy:1992}, in which a spin-1 particle is formed out of two spin-1/2 particles. 
Thus, spin-1/2 ladder systems emerge as an experimentally realisable platform for the Haldane phase. 
While a natural implementation with spin-1 particles on individual rungs requires ferromagnetic rung couplings and antiferromagnetic leg couplings, a much wider variety of couplings in spin-1/2 quantum ladders features the Haldane phase \cite{Hida1992,white:1996}. This includes the antiferromagnetic Heisenberg case, which we realise here as the strong-interaction limit of the Fermi-Hubbard model.


\begin{figure}[t]
\centering

\includegraphics{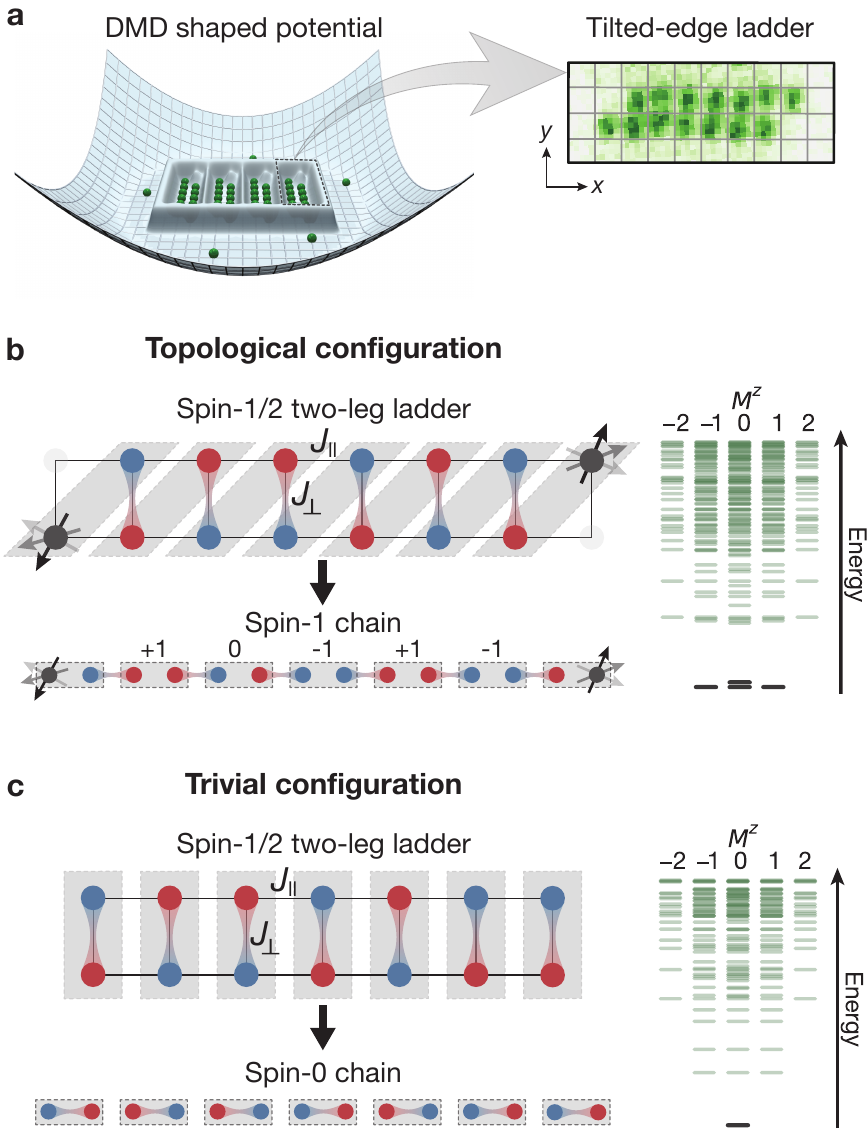}
\caption{\textbf{Probing topological phases in spin-1/2 ladders of cold atoms.} 
\textbf{a}, Realisation of tailored spin-1/2 ladders in a single plane of a 3D optical lattice with a potential shaped by a digital micromirror device (DMD). 
The dilute wings of the potential are well separated from the homogeneous ladder systems. 
Using quantum gas microscopy, we obtain fully spin- and density-resolved images of the system. 
The inset shows a single-shot fluorescence image of the prepared ladder without spin resolution. 
\textbf{b} and \textbf{c}, Connecting spin-1/2 ladders to integer-spin chains by grouping pairs of spins in unit cells. For diagonal unit cells (\textbf{b}) the AFM Heisenberg ladder adiabatically connects to the Haldane spin-1 chain showing spin-1/2 edge states and hidden long-range order (i.e. AFM order interspersed with $S^z=0$ unit cells). We realise this topological configuration by blocking one site on each end of the ladder. For the case of straight edges (\textbf{c}), the unit cells coincide with rungs of the ladder and the system is in the topologically trivial configuration. For $J_\parallel \ll \Jp$, singlets form on the rungs leading to a spin-0 chain. 
The energy spectra of the systems grouped by total magnetisation $M^z$ display gapped fourfold near-degenerate ground states for the topological configuration and a single ground state for the trivial one. Sketch for $L=7$.
}
\label{fig:fig1}
\end{figure}

In our experiment, we prepare such ladders by adiabatically loading a spin-balanced mixture of the two lowest hyperfine states of $^6$Li into an engineered lattice potential (see Methods).
As illustrated in Fig.~\ref{fig:fig1}a, we realise four isolated two-leg ladders with variable number of $L$ unit cells, surrounded by a low-density bath of particles\cite{mazurenko:2017}.
The atoms in the lowest band of the optical lattice are well described by the Fermi-Hubbard model with tunnelling energies $t_\parallel$ (chain), $t_{\perp}$ (rung) and on-site interactions $U$. For half-filling and at strong $U/t_{\parallel,\perp}\sim13$, used throughout most of our experiments (see Methods for details), density fluctuations are suppressed and the spin ladder realises the Heisenberg model\cite{auerbach:1994}
\begin{align}\label{eq:eq1}
\hat{H} = \Jpl \sum_{\substack{x\in[0,L)\\y=A,B}} \hat{\mathbf{S}}_{x,y} \cdot \hat{\mathbf{S}}_{x+1,y} + \Jp \sum_{x\in[0,L)} \hat{\mathbf{S}}_{x,A}\cdot\hat{\mathbf{S}}_{x,B}
\end{align}
with positive leg and rung couplings, $J_{\parallel,\perp}=4t_{\parallel,\perp}^2/U$ and the spin-1/2 operators $\hat{\mathbf{S}}_{x,y}$ at site $(x,y)$ with $A,B$ denoting the two legs of the ladder.

The topological properties are most easily explained in the limit $\Jp\gg \Jpl$, where strong spin singlets form along the rungs and the system exhibits an energy gap of $\Jp$. 
The behaviour on the edges of the ladder then depends on how the system is terminated. For tilted edges (see Fig.~\ref{fig:fig1}b), two unpaired spin-1/2s  remain and the many-body system has a fourfold degeneracy that is only weakly lifted by an edge-to-edge coupling which vanishes exponentially with system size (see SI). 
In the trivial case of straight edges (see Fig.~\ref{fig:fig1}c), all spins pair into singlets and the ground state is unique. 
These descriptions remain valid even for weaker $\Jp/\Jpl$, where the singlet alignment may change between vertical and horizontal, but any line between two rungs cuts an even number of singlets \cite{Kim2000,Bonesteel1989}.   

To make the analogy between the spin-1/2 ladder and the Haldane integer chain more apparent, we switch to a description in terms of total spin per $k$th unit cell, ${\hat{\mathbf{S}}}_k = {\hat{\mathbf{S}}}_{k,A} + {\hat{\mathbf{S}}}_{k,B}$, where the indices (A,B) indicate the two spin-1/2s in the same unit cell making ${\hat{\mathbf{S}}}_k$ an integer spin. 
In the diagonal unit cell such a system shows a high ($\geq80\%$) triplet fraction \cite{white:1996} (see SI).
We note that this spin ladder can be adiabatically connected to a spin-$1$ chain by including ferromagnetic couplings within the unit cell \cite{Hida1992}. 
However, having a high triplet fraction is not essential for having a well-defined Haldane phase as both systems share the same universal SPT features \cite{white:1996}.


\begin{figure}[t]
	\centering
	\includegraphics{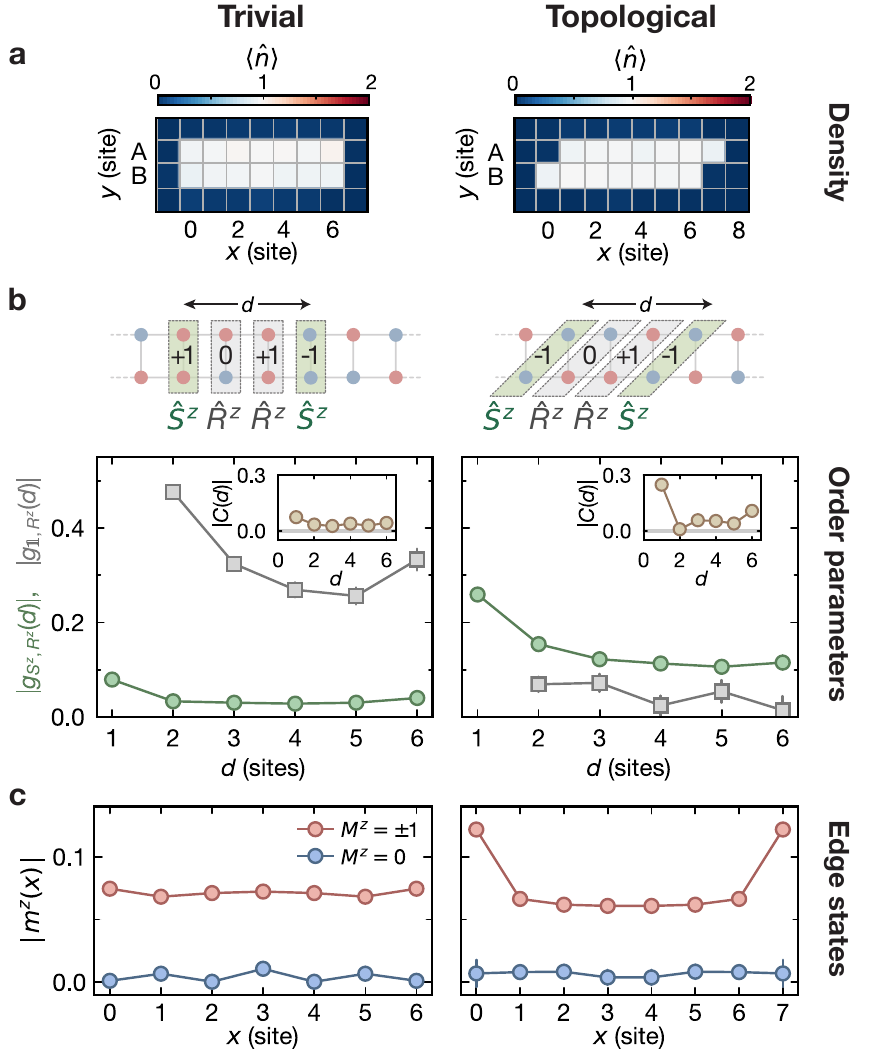}
	\caption{
		\textbf{Trivial versus topological configurations.} 
		\textbf{a}, The atomic density distribution $\langle \hat{n} \rangle$ of straight- and tilted-edge ladders.
		\textbf{b}, The amplitudes of the spin-string correlator $\gS$ (green circles) and the string-only correlator $\gI$ (grey squares) observed as a function of the spin distance over $d$ unit cells. The cartoon illustrates the unit cells, the spin total spin $S^z$ per unit cell, and the string correlators for a subsystem with $d=3$.		 
		In the trivial configuration (rung unit cells), $\left|\gI(d)\right|$ is well above zero, whereas $\left|\gS(d)\right|$ is rapidly vanishing at $d>1$. 
		In contrast, for the topological configuration (diagonal unit cells), $\left|\gS(d)\right|$ shows a long-range correlation, while $\left|\gI(d)\right|$ is close to zero.
		In both cases, the two-point spin-spin correlation $C(d)$ decays rapidly to zero as a function of the distance $d$ (insets). 
		The correlators $\gI,\gS$ and $C(d)$ are evaluated for fixed total magnetisation $M^z = 0$. 
		\textbf{c}, Amplitudes of the rung- and inversion-averaged local magnetisations $\left|m^z(x)\right|$ plotted as a function of position $x$ along the chains for different $M^z$. In the imbalanced spin sector of the topological configuration ($M^z=\pm1$), the result displays a localisation of the excess spins at the edges signalling the presence of edge states. All data was taken with $\Jp/\Jpl=1.3(2)$. Error bars denote one standard error of the mean (s.e.m) and are smaller than their marker size if not visible.
	}
	\label{fig:fig2}
\end{figure}
The defining property of the Haldane SPT phase is that it is an integer spin chain with spin-$1/2$ edge modes: the bulk $SO(3)$ symmetry is said to \emph{fractionalise} into $SU(2)$ symmetry at the edge.
It has no spontaneous symmetry breaking and thus the spin correlation function $\langle \hat{S}^z_k \hat{S}^z_{k+d}\rangle$ is short-ranged.
Instead, the aforementioned symmetry fractionalisation\cite{pollmann:2010,chen:2011c} can be detected in the bulk using string order parameters \cite{dennijs:1989,pollmann:2012a}
\begin{equation}\label{eq:eq2}
g_{\mathcal{O},U}(d) = \left\langle \hat{\mathcal{O}}_k\left( \prod_{l = k+1}^{k+d-1} \hat U_l \right) \hat{\mathcal{O}}_{k+d} \right\rangle
\end{equation}
with an on-site symmetry $\hat U_l$ and endpoint operator $\hat {\mathcal{O}}_k$ where $l$ denotes the unit cell and $d$ the string distance (see Fig.~\ref{fig:fig2} and SI).
This correlator probes the transformation behaviour of the bulk under a symmetry $\hat{U}_l$, e.g. a spin rotation around the $z$-axis by $\pi$, $\hat{R}^z_l \equiv \exp \big( i \pi \hat S^z_l \big)$. 
The pure-string correlator $g_{\mathds{1}, {R^z}}(d)$, where $\hat {\mathcal{O}}_k=\mathds{1}$ and $\hat{U_l} = \hat{R}^z_l$, is non-zero for $d\gg1$ if the edge does not have half-integer spins \cite{pollmann:2012a}. This is the case for the topologically trivial configuration but not for the Haldane phase where the symmetry is fractionalised. The spin-string operator $g_{S^z,R^z}(d)$ \cite{dennijs:1989}, $\hat {\mathcal{O}}_k=\hat{S}^z_k$, exhibits the opposite behaviour and is non-zero only in the Haldane phase (see SI for details about the symmetries of the Haldane phase). Thus we can identify the Haldane phase by comparing the two string correlators \gS and \gI and observe opposite behaviour in the two different regimes..

Another perspective on \gS can be gained by recognising it as a normal two-point correlator at distance $d$ which ignores all spin-0 contributions along the way (``squeezed space'' \cite{ogata:1990,hilker:2017}). In the underlying spin-1/2 system, this order stems from $N-1$ consecutive rungs dominantly consisting of $N-1$ singlets and two spin-1/2s which have a combined total spin of +1, 0, or $-$1. 



In order to observe the characteristics of the SPT phase, we prepare a two-leg ladder of length $L=7$ and $\Jp/\Jpl=1.3(2)$ in both the topological and the trivial configuration. 
The tailored potential yields a homogeneous filling of the system with sharp boundaries (see Fig.~\ref{fig:fig2}a) characterised by a remaining density variance over the system of $2 \times 10^{-4}$. 
To focus on the spin physics, we select realisations with $N_{\uparrow}+N_{\downarrow}=2L$ per ladder. Additionally, we exclude ladders with an excessive number of doublon-hole fluctuations and do not consider strings with odd atom numbers in the string or the endpoints of the correlator (see Methods). 
We characterise the spin-balanced ladder systems ($M^z\equiv(N_{\uparrow}-N_{\downarrow})/2=0$) by evaluation of the string order parameters as defined in Eq.~(\ref{eq:eq2}). 
In the topological configuration, we observe a fast decay of $\gS$ over a distance of $\sim$1 site and a long-range correlation up to $d=6$ with a final value of $\gS\simeq0.1$ (see Fig.~\ref{fig:fig2}b).
In contrast, for the trivial configuration, the correlation decays rapidly to zero as a function of the string correlator length.
The opposite behaviour is seen for $\gI(d)$, demonstrating the hidden correlations expected for both phases. 

Furthermore, the two--point spin correlation, $C(d)\equiv g_{S^z,\mathds{1}}(d)=\langle \hat{S}^z_{k}\hat{S}^z_{k+d}\rangle$, yields only the short--range AFM correlation characteristic for a gapped phase (see insets).
It is interesting to note that at the largest distance in the topological case, $C(d=6)$ displays a clear (negative) correlation between the two edge spins, despite small correlations at shorter distances. This (classical) correlation confirms the existence of a non-magnetised bulk, such that spins on the edges of the system must be of opposite direction at global $M^z = 0$.

\begin{figure*}[t]
\centering
\includegraphics{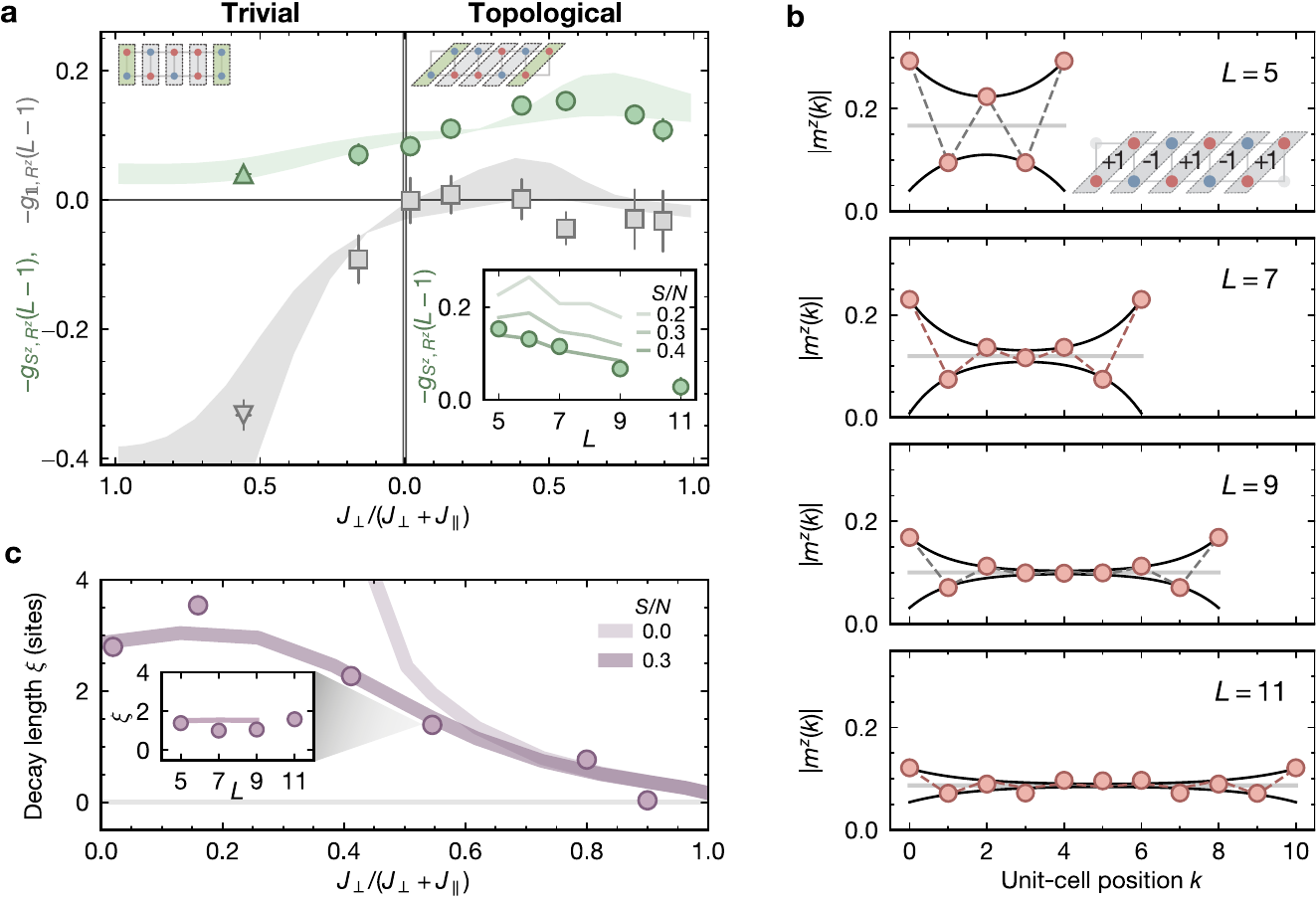}
\caption{\textbf{Influence of spin-coupling strength on the string order parameters and the edge states.} 
\textbf{a}, The two string order parameters, $\gS$ (green) and $\gI$ (grey), for both trivial (left) and topological (right) configurations measured as a function of $\Jp/(\Jp+\Jpl)$. 
Both $\gS$ and $\gI$ stay finite in their respective phases and are largely consistent with zero in the other phase. The data was taken at a chain length of $L=5$ except one data point marked by a triangle at $L=7$.
Shaded curves are the exact diagonalisation (ED) results of the two order parameters at finite entropy-per-particle, $S/N =(0.3-0.45)\,\kB$ and $L=5$. 
The inset shows the measured $\gS$ as a function of the chain length $L$ at $\Jp/\Jpl=1.3(2)$ (i.e. $\Jp/(\Jp+\Jpl)= 0.56(4)$). The decay in the magnitude of the string order parameter with length is expected at finite temperatures in quantitative agreement with ED results (lines) at $S/N\approx 0.4\,\kB$.
\textbf{b}, Edge state localisation at $\Jp/\Jpl=1.3(2)$. In the $M^z=\pm1$ spin sectors of the topological configuration, the unit-cell local magnetisation $|m^z(k)|$ at chain position $k$ shows excess magnetisation localised at the edges for different lengths. The black line is a fit to our inversion-averaged data.
\textbf{c}, The localisation length $\xi$ of the edge states increases with the leg coupling $\Jpl$ but saturates at a value set by temperature and system size $L=5$. Lines are ED results at $S/N = 0.3\,\kB$ and $0\,\kB$.
The inset shows the independence of $\xi$ with respect to $L$ extracted from the plots in \textbf{b} as well as ED results for $S/N=0.3\,\kB$. Error bars denote one standard error of the mean (s.e.m) and are smaller than their marker size if not visible.
}
\label{fig:fig3}
\end{figure*}

We probe the edges explicitly by measuring the amplitude of the local rung--averaged magnetisation $m^z(x)$ as a function of rung position $x$ for different sectors of the ladder magnetisation $M^z$ (see Fig.~\ref{fig:fig2}c). In the case of an imbalanced spin mixture with $M^z=\pm1$, we see that the two end sites exhibit on average a higher magnetisation than the bulk rungs in the topological configuration. 
This is consistent with the bulk of the ground states of both phases forming a global singlet and only the edges of the topological phase carrying an excess spin-1/2 without energy cost. The measured bulk magnetisation can be attributed to finite temperature effects (see SI).

The SPT phase is expected to be robust\cite{white:1996} upon variation of the ratio $\Jp/\Jpl$, but maintains a finite gap in the system.
We realise both the trivial and topological configuration with different $t_{\perp}/t_\parallel$ at almost fixed $U$ and study the string correlators at maximal distance ($L-1$) versus $\Jp/\Jpl$ (see Fig.~\ref{fig:fig3}a). 
For the topological configuration, we observe $\gI(L-1)\simeq0$ and $|\gS|>0$ for all $\Jp/\Jpl$ with a maximum around $\Jp/\Jpl \simeq 1.3(2)$ (i.e. $\Jp/(\Jp+\Jpl)\simeq 0.56(4)$), while for the trivial case, the role of the correlators is reversed. 
Both phases continuously connect in the limit of two disconnected chains at $\Jp=0$.
These observations demonstrate qualitatively all the key predictions of the antiferromagnetic spin-1/2 ladder at $T=0$\cite{white:1996} and the strength of the measured correlations are consistent with exact diagonalisation (ED) calculations using an entropy per particle $S/N = (0.3-0.45)\,\kB$ (shaded lines in Fig.~\ref{fig:fig3}a). 

We reveal these features despite the finite temperature in our system, which would destroy the long-range hidden order in an infinite system. The total entropy in our system is, however, still low enough to yield a large fraction of realisations of the topological ground state. In larger systems, the total number of thermal excitations grows (at fixed entropy per particle) and the non-local correlator $|\gS(L-1)|$ decreases (see inset of Fig.~\ref{fig:fig3}a), consistent with vanishing correlations in the thermodynamic limit thus yielding a restriction on our system size at our level of experimental precision and entropy per particle (see SI). Finite size effects are explored in detail in the SI. 
We note that the difference between the SPT phase and the trivial phase is here clearly shown by considering both \gS and \gI.

To investigate the localisation length of the edge states, we evaluate our data for $M^z=\pm1$ and plot the local magnetisation per unit cell $m^z(k)$ for different system sizes (see Fig.~\ref{fig:fig3}b). 
Due to the singlets in the bulk, the excess spin is most likely found at the edges of the system. This spin partly polarises the neighbouring sites antiferromagnetically leading to an exponentially localised net magnetisation with AFM substructure \cite{miyashita:1993}.
The data is well described by the fit function $m^z(k)= m_B+m_E\left((-1)^ke^{-k/\xi}+(-1)^{L-k-1}e^{-(L-k-1)/\xi}\right)$ with free bulk magnetisation $m_B$, edge magnetisation $m_E$, and decay length $\xi$. In Fig.~\ref{fig:fig3}c, we show how this localisation length $\xi$ decreases as we approach the limit of rung singlets $\Jp\gg\Jpl$. Comparison with ED lets us identify two regimes: at $\Jp\gtrsim\Jpl$, the measured $\xi$ drops with larger $\Jp$ and coincides with theory independent of temperature, while at low $\Jpl$ thermal effects dominate, limiting the increase of $\xi$ to 3 sites for our system (see SI).

\begin{figure}[t]
\centering
 \includegraphics{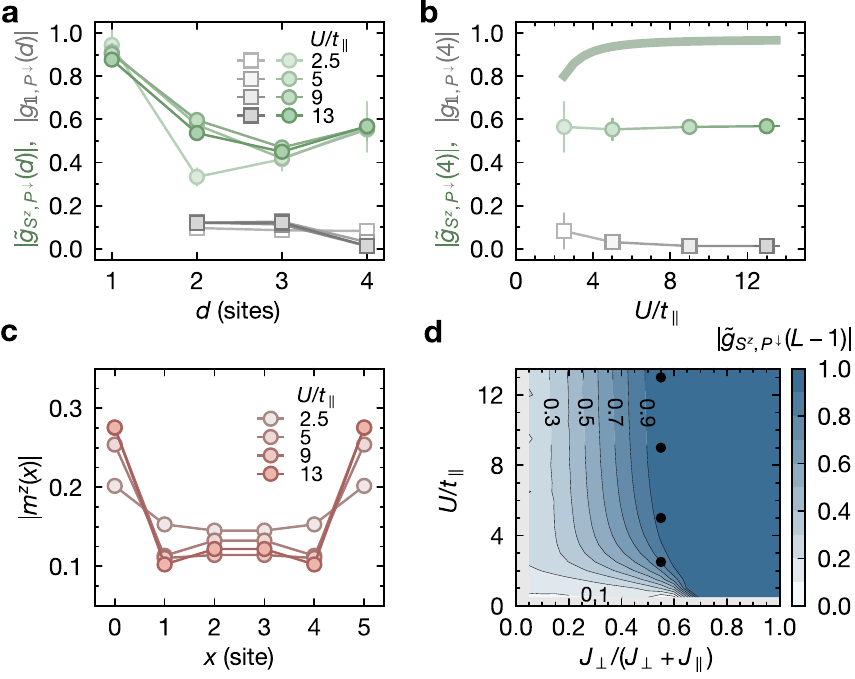}
\caption{\textbf{Robustness of the Haldane phase to density fluctuations.} 
\textbf{a}, \textbf{b}, The hidden SPT order is preserved even at low Hubbard interactions as revealed by the novel string correlators $\left|\gtP(d)\right|$ (green circles) and $\left|\gIP(d)\right|$ (grey squares) based on the spin-down parity $P^{\downarrow}$. $\left|\gtP\right|$ stays non-zero while $\left|\gIP\right|$ is consistent with zero for $d=L-1$ over the measured interaction range. 
The same qualitative behaviour is seen in zero temperature DMRG calculations (shaded line) with $L\rightarrow \infty$.
\textbf{c}, Spatial distribution of excess magnetisation ($M^z=\pm1$) for decreasing $U/t_{\parallel}$. Even far away from the Heisenberg regime, the edge state signal remains strong and only diminishes for very weak $U/t_{\parallel}$. 
\textbf{d}, Map of zero temperature DMRG ($L\rightarrow \infty$) results for the spin string correlator in the entire parameter space of the topological phase. It shows a strictly non-zero \gSzP while $\gIP(L-1)=0$ everywhere in this phase.
The black circles indicate the parameters of the measurements. All experimental data were taken at $\Jp/\Jpl=1.3(2)$ and $L=5$ in the tilted geometry. $M^z=0$ in a, b, and d. Error bars denote one standard error of the mean (s.e.m) and are smaller than their marker size if not visible.
}
\label{fig:fig4}
\end{figure}

Thus far, we have worked in the Mott limit where density fluctuations can be ignored, such that the spin Hamiltonian Eq.~(\ref{eq:eq1}) is a good effective description of the Fermi-Hubbard ladder. 
However, it is known that the Haldane SPT phase can be unstable to density fluctuations \cite{anfuso:2007,moudgalya:2015,verresen:2021}.
By reducing $U/t_\parallel$, the symmetry in the unit cell in the bulk changes from $SO(3)$ to $SU(2)$ as it now may contain both half-integer and integer total spin. This effectively removes the distinction between bulk and edge (see SI).
This means that the edge mode and string order parameter are no longer topologically non-trivial, which is also manifested in the fact that the two phases can be adiabatically connected by tuning through a low-$U/t_\parallel$ regime if one breaks additional symmetries but preserves spin-rotation symmetry \cite{anfuso:2007,moudgalya:2015,verresen:2021}. In particular, the above string orders lose their distinguishing power: $g_{S^z,R^z}$ and $g_{\mathds{1}, {R^z}}$ will \emph{both} generically have long-range order away from the Mott limit \cite{anfuso:2007}.

Intriguingly, despite the breakdown of the above symmetry argument and string order parameter, the Hubbard ladder (with diagonal unit cell) remains a non-trivial SPT phase due to its sublattice symmetry. This symmetry is a direct consequence of the ladder being bipartite (see SI for details). It is simplest to see that this protects the SPT phase in the limit $U=0$, where the two spin species decouple, such that our model reduces to two copies of the SSH chain \cite{su:1979}. It is known that such a stack remains in a non-trivial SPT phase in the presence of interactions, i.e. $U \neq 0$ \cite{verresen:2017}. 
Moreover, together with the parity symmetry of spin-down particles, $\hat P^{\downarrow}_l \equiv \exp\left[ i \pi \left(\hat n^{\downarrow}_{l,A} + \hat n^{\downarrow}_{l,B} \right)\right]$, it then gives rise to a different string order parameter: the topological phase is characterised by long-range order in $g_{S^z,P^\downarrow}$ whereas it has vanishing correlations for $g_{\mathds{1},P^\downarrow}$, with the roles being reversed in the trivial phase. This novel string order is derived in the SI. Remarkably, in the Heisenberg limit, it coincides with the conventional string order parameter used before. 

In the topological phase it is meaningful to normalise \gSzP~ to $\gtP = \eta ~ \gSzP$ with $\eta^{-1}=\left\langle \big|\hat{S}^z_k\big|\big|\hat{S}^z_{k+d}\big|\right\rangle$, which effectively excludes endpoints with spin $S^z=0$. We indeed find unchanged string correlations \gtP~ and \gIP~ down to the lowest experimentally explored value $U/t_\parallel = 2.5(2)$ (see Fig.~\ref{fig:fig4}a,b) and edge state signals down to $U/t_\parallel = 5.0(3)$ (see Fig.~\ref{fig:fig4}c). DMRG calculations for $L\rightarrow \infty$ confirm non-zero \gtP$(L-1)$ at $T=0$ and for all rung coupling strengths (see Fig.~\ref{fig:fig4}d), while $\gIP(L-1)$ is strictly zero. Due to the normalisation \gtP$(L-1)$ goes to 1 for $\Jp\gg\Jpl$.

In our work, we realised a finite temperature version of the topological Haldane SPT phase using the full spin and density resolution of our Fermi quantum gas microscope. We demonstrated the robustness of the edge states and the hidden order of this SPT phase in both the Heisenberg and the Fermi-Hubbard regime. 
In the future, studies may extend the two-leg ladder to a varying number of legs where one expects clear differences between even and odd numbers of legs \cite{schulz:1986} and topological effects away from half-filling \cite{nourse:2016}, or investigate topological phases in higher dimensions \cite{szasz:2020}. Furthermore, the ladder geometry holds the potential to reveal hole-hole pairing\cite{noack:1995} at temperatures more favourable than in a full 2d system.

\bigskip

\textbf{Acknowledgments:} This work was supported by the Max Planck Society (MPG), the European Union (FET-Flag 817482, PASQUANS), the Max Planck Harvard Research Center for Quantum Optics (MPHQ), the Cluster of Excellence 'CUI: Advanced Imaging of Matter' of the Deutsche Forschungsgemeinschaft (DFG) - EXC 2056 - project ID 390715994 and under Germany's Excellence Strategy -- EXC-2111-390814868. J.K. acknowledges funding from Hector Fellow Academy and T.C. from the Alexander v. Humboldt foundation. R.V. is supported by the Harvard Quantum Initiative Postdoctoral Fellowship in Science and Engineering and by the Simons Collaboration on Ultra-Quantum Matter (Simons Foundation, 651440, Ashvin Vishwanath). F.P. acknowledges the support of the European Research Council (ERC) under the European Unions Horizon 2020 research and innovation program (grant agreement No. 771537).

\textbf{Author contributions}: P.S, S.H and D.B planned the experiment and analysed the data. P.S, S.H, D.B and T.C contributed significantly to the data collection and ED calculations. J.B., R.V., F.P performed the DMRG simulations and analytical calculations. T.A.H., C.G. and I.B. supervised the study.  All authors contributed extensively to interpretation of the data and production of the manuscript.

\textbf{Competing interests}: The authors declare no competing interests.

\textbf{Data availability}: The datasets generated and analysed during the current study are available from the corresponding author on reasonable request.
\newpage
\bibliographystyle{naturemag}

\clearpage
\newpage
\makeatletter 
\renewcommand{\thefigure}{S\@arabic\c@figure}
\makeatother
\setcounter{figure}{0}
\setcounter{table}{0}
\section*{Supplementary Information}
\subsection*{Experimental sequence}
In each experimental run, we prepare a cold atomic cloud of $^6$Li in a balanced mixture of the lowest two hyperfine states ($F=1/2, m_F = \pm 1/2$). 
For evaporation, we confine the cloud in a single layer of a staggered optical superlattice along the $z$-direction with spacings $a_\mathrm{s} = \SI{3}{\micro\meter}$ and $a_\mathrm{l} = \SI{6}{\micro\meter}$ and depths $V_s = 45\, E_\mathrm{R}^\mathrm{s}$ and $V_l = 110\,E_\mathrm{R}^\mathrm{l}$, where $E_\mathrm{R}$ denotes the recoil energy of the respective lattice. The atoms are harmonically confined in the  $xy-$plane and the evaporation is performed by ramping up a magnetic gradient along the $y$-direction (see~\cite{koepsell:2020}). The final atom number is tuned via the evaporation parameters. 

The cloud is then loaded into an optical lattice in the $xy-$plane with spacings  $a_x = \SI{1.18}{\micro\meter}$ and $ a_y = \SI{1.15}{\micro\meter}$, which is ramped up within $\SI{100}{\milli\second}$ to its final value ranging from $5\,E_\mathrm{R}$ to $15\,E_\mathrm{R}$ depending on the chosen Hubbard parameters. 
The scattering length is tuned from $230\,a_\text{B}$ during evaporation, with $a_\text{B}$ being the Bohr radius, to its final value ranging between $241 \, a_\text{B}$ and $1200\,a_\text{B}$, using the broad Feshbach resonance of $^6$Li.
An overview of the parameters of each dataset is given in the Table \ref{tab:params}. 
Simultaneously with the lattice loading, a repulsive potential is ramped up which compensates for the harmonic confinement generated by the curvature of the Gaussian lattice beams and divides the resulting flat area into four disconnected ladder systems along the $y-$direction (see below ``potential shaping''). 
We achieve temperatures of $\kB T \sim 0.9(3)\,\Jpl$ for the parameters in Fig.~\ref{fig:fig2}.

For detection, the configuration is frozen by ramping the $xy-$lattices to $43 \,E_\mathrm{R}^{xy}$ within $\SI{250}{\micro\second}$. A Stern-Gerlach sequence separates the two spin species into two neighbouring planes of the vertical superlattice, which are then separated to a distance of $\SI{21}{\micro\meter}$ using the charge pumping technique described in \cite{koepsell:2020}. 
Finally, simultaneous fluorescence images of the two planes are taken using Raman sideband cooling in our dedicated pinning lattice with an imaging time of $\SI{2.5}{\second}$ \cite{omran:2015}.
The fluorescence of both planes is collected through the same high-resolution objective. The light is then split into two paths using a polarizing beam splitter. One of the beams passes through a variable 1:1 telescope before both paths are recombined on a second polarizing beamsplitter with a small spatial offset. This allows us to image both planes in a single exposure, each plane in focus on a separate fixed position of our camera. We calibrated the relative position on the camera of the two imaged planes using 300 shots of a spin-split Mott insulator and the matching algorithm described in the supplement of \cite{koepsell:2020}. 
The overall detection fidelity per atom is $\SI{96(1)}{\percent}$.
\begin{table}[t]
	\begin{tabular}{| c | c | c | c | c | }
		\hline
		\textbf{length}     &  \textbf{hopping }  &\textbf{hopping }  & \textbf{interaction}    & \textbf{ratio}            \\ 
		$L$ (sites) &  $t_\parallel /h (\SI{}{\hertz})$ & $t_\perp/h (\SI{}{\hertz})$ & $U/h  (\SI{}{\hertz})$ & $J_\perp/J_\parallel$ \\ \hline
		5,6,7,9,11 & 250 & 280 & 3500 & 1.3(2) \\ \hline
		5          & 330 & 38  & 4000 & 0.013(2)    \\
		5          & 300 & 130 & 3600 & 0.20(3)              \\
		5          & 340 & 280 & 3000 & 0.7(1)               \\
		5          & 250 & 280 & 3500 & 1.3(2)             \\
		5          & 150 & 300 & 3500 & 4.0(6)                 \\
		5          & 130 & 390 & 3300 & 8(1)                 \\ \hline
		5          & 250 & 280 & 3500 & 1.3(2)    \\
		5          & 250 & 280 & 2300 & 1.3(2)             \\
		5          & 250 & 280 & 1250 & 1.3(2)              \\
		5          & 250 & 280 & 650  & 1.3(2)          \\ \hline   
	\end{tabular}
	\caption{\textbf{Experimental parameters.} The parameters system size $L$, leg coupling $t_\parallel$, rung coupling $t_\perp$, interaction $U$ and the resulting ratio $J_\perp/J_\parallel$ are shown for all datasets. The uncertainties are given for $\Jp/\Jpl$ and originate from a 5\% uncertainty on the hopping parameters $t_\perp$ and $t_\parallel$. For the length scan we keep all other parameters constant, whereas the $J_\perp/J_\parallel$ scan demands a tuning of both tunnelling amplitudes in order to keep both $U/t_\parallel$ and $U/t_\perp$ high. Where the topologically trivial geometry is realized, it has the same parameters as the topological geometry.}
	\label{tab:params}
\end{table}

\subsection*{Potential shaping}
\begin{figure}[t]
	\centering
	\includegraphics{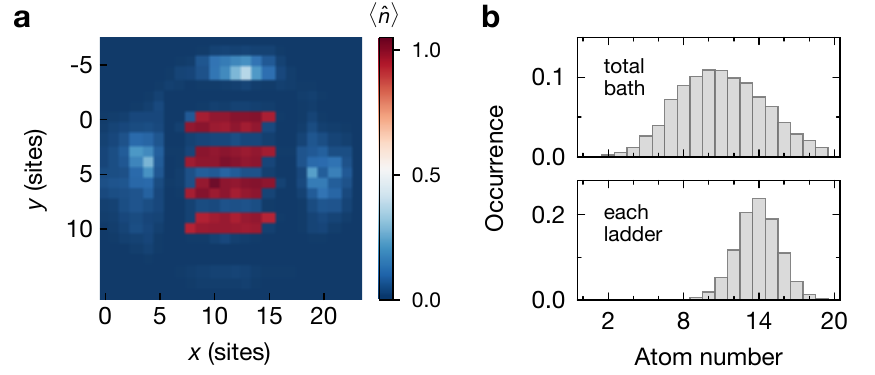}
	\caption{\textbf{Density engineering.} \textbf{a}, Repulsive light shaped with a DMD splits the system into 4 independent ladders in the centre surrounded by a low-density bath. The density of the ladders is $n=0.992$ with a standard deviation of 0.03. \textbf{b}, The occupation  histograms show the normalised occurrence of total atom numbers in each ladder and the normalised occurrence in the surrounding bath for $L=7$. 
  Almost 25\% of the ladder realisations have $N=2L$. }
	\label{fig:S2}
\end{figure}

The ladder systems are created by superimposing the optical lattice with a repulsive potential, which is shaped by projecting incoherent light at $\SI{650}{\nano\meter}$ (generated from a SLED by Exalos EXS210030-03) from a digital micromirror device (DMD) through the high-resolution objective.
Four ladders are created by blocking lattice sites with a potential  $V=3.5(5),U$ around each ladder. 
The area outside of the walled ladders is lifted above the inner ladder potential but remains below the interaction energy $U$. It thus serves as a reservoir for surplus atoms, which occupy this region once the lowest Hubbard band of the ladders is filled. 
The flatness of the potential is adjusted for each parameter setting since the intensity of the lattice beams directly influences the curvature of the potential. 
This is accomplished by realizing a system with about 20\% doping and returning the average density of 100-150 experimental runs as feedback to the DMD pattern. We repeat the feedback until we reach a sufficiently flat density profile with a variance $<1\times10^{-3}$ over the $8L$ ladder sites. 
To adjust for drifts in the lattice phase, we continuously track the lattice phase of each experimental run and feedback to the potential position accordingly. 
In Fig. \ref{fig:S2}, the average density and occupation histograms of all four ladders and the reservoir area are shown for the dataset of $L=7$. 

\subsection*{Data selection}
In each experimental run, four ladder systems are realised. 
To fulfil the criteria of the Heisenberg regime, we then select on ladder instances with atom number $N=2L$ and restrict the total magnetisation to $M^z =0$, $|M^z| = 1$, or  $|M^z| \leq 1$, depending on the observable, and specify the magnetisation sector whenever data points are presented. $|M^z|\leq1$ for \SI{87.5}{\percent} of all data. Additionally, for all measurements in the Heisenberg regime, we remove ladders with more than two doublons as those indicate a mismatch of the DMD pattern relative to the lattice phase.
To give a specific example, we here give the precise numbers for the data presented in Fig.~\ref{fig:fig2}. This dataset consists of 7533 realisations with four ladders each. Out of those 28128 ladders, 6721 have an atom number of 14. 2636 ladders then have a total magnetisation $M^z=0$, 3094 have a magnetisation fo $M^z = \pm 1$.  Finally of those 2636, 77 have more than 2 doublon-hole pairs, which we exclude as those are most likely caused by drifts of the potential pattern given by the DMD. This leaves then in total 2559 ladders out of the initial 28128 for the calculation of the string correlator.

For calculating the string correlators $\gS$ and $\gI$ at fixed $d$, we exclude realisations with odd atom number in the bulk area (grey area in the cartoon of Fig.~\ref{fig:fig2}b) as those would lead to imaginary contributions to the correlators and also exclude odd atom number at the edge areas (green in the cartoon of Fig.~\ref{fig:fig2}b). These cases are mostly due to the finite $U/t_\parallel$ which still allows for some particle fluctuations. We keep other particle-hole fluctuations like those occurring within the string. Those do not alter the observed string correlation relative to the Heisenberg model.

\subsection*{Nearest-neighbour spin correlations}
\begin{figure}[t]
	\centering
	\includegraphics{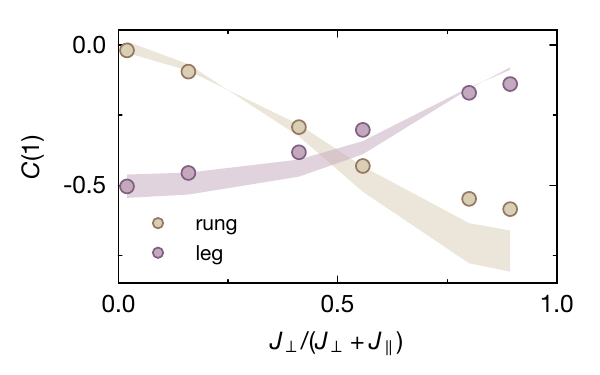}
	\caption{\textbf{Nearest-neighbour spin correlations.} The nearest-neighbour spin correlation $C(1)$ for different $J_{\perp}/J_\parallel$ in the $L= 5$ system. The brown (purple) points refer to the correlations along the rung (leg). The shaded areas correspond to the correlations in the Heisenberg model with an entropy of $S/N = (0.3-0.4)\,\kB$ per particle. Both theoretical and experimental values are obtained from the magnetisation sector $M^z = 0$.}
	\label{fig:S3}
\end{figure}

The two-leg ladder systems show strong antiferromagnetic spin correlations whose dominant orientation depends on the ratio of couplings $J_{\perp}/J_\parallel$ and whose strength is measured by $C_x(d) = 4\langle \hat{S}^z_{i,j}\hat{S}^z_{i,j+d}\rangle$ and $C_y = 4\langle \hat{S}^z_{A,j}\hat{S}^z_{B,j}\rangle$.
For a leg coupling $J_\parallel$ much higher than the rung coupling $J_{\perp}$, the nearest-neighbour spin correlator $C_y$ along the rung almost vanishes, whereas correlations reach $C_x(1) = -0.500(6)$ along the leg direction. 
For a dominating rung coupling $J_{\perp}$, the $C_y$ reaches $-0.58(1)$, indicating singlet formation between the two sites of a rung. 
Fig.~\ref{fig:S3}a shows the measured spin correlations along both rung and leg for different values of $J_{\perp}/J_\parallel$. The values match the finite temperature Heisenberg model for an entropy of $S/N = (0.3-0.4)\,\kB$ per particle obtained from ED. 

\subsection*{Theory simulation}
In this work, we have employed two different numerical methods to obtain theoretical predictions for the experimentally measured observables. The results in the Heisenberg regime were obtained using exact diagonalisation (ED) of our spin-1/2 ladders up to sizes of $L=9$ (limited by computational resources). For each data point, the system size and geometry in the ED simulation is the same as in the experimental data. The finite temperature results were obtained by using the full spectrum. We specify the entropy per particle $s=S/N$, which we find to be approximately independent of coupling parameters in the experimental realisations. 
The results in the Hubbard regime are calculated using the Density Matrix Renormalisation Group (DMRG) Ansatz~\cite{white:1992} based on the TeNPy library (version 0.3.0)~\cite{hauschild:2018}. For all calculations, we conserved the total particle number and the total magnetisation. For the phase diagram in Fig.~\ref{fig:fig4}d we used the iDMRG technique to obtain the ground state and the values of the string order parameters in the thermodynamic limit.  For this, we evaluated the ground state for each parameter and used a maximal MPS bond dimension $\chi=1200$. The bond dimension is increased in steps $\Delta\chi=40$ and the simulation stopped when the difference in the ground state energy per unit cell $E(\chi+\Delta\chi)-E(\chi)<10^{-7}$. This worked for most parameters except in the vicinity of two decoupled Hubbard chains and at small values of $U/t_\parallel$. Nevertheless, in this regime we find that the energy per unit cell is converged up to $E(1200)-E(1160)<10^{-4}.$ For the experimentally accessible regime all calculations fulfil the former bound. To obtain the infinite length value of the string order parameters, we calculated it for different lengths $d\in [200,400,...,1600]$ to make sure that its final value is converged.

\subsection*{Triplet fraction in the unit-cell}
\begin{figure}[t]
	\centering
	\includegraphics{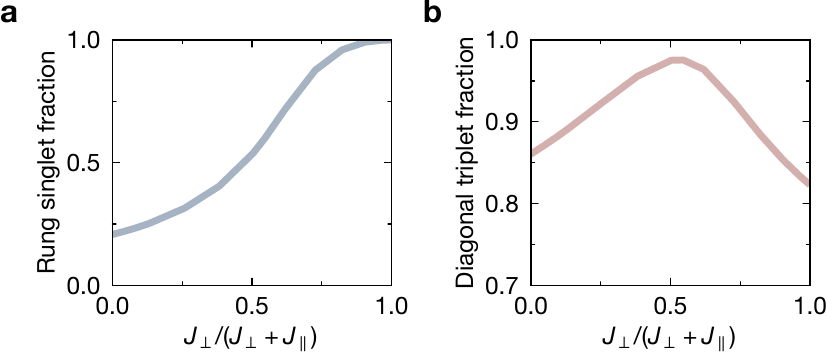}
	\caption{\textbf{Singlet and triplet fractions.} \textbf{a}, Numerical singlet fraction on the rung of the ladders for different $\Jp/\Jpl$ at $L=5$, $M^z = 0$. The singlet fraction increases monotonically with the rung coupling. \textbf{b}, The fraction of triplets in the diagonal unit-cell is higher than 80\% for all $\Jp/\Jpl$ and peaks close to 1 when rung and leg coupling become comparable at $\Jp/\Jpl \sim 1$. Both plots are calculated on ladders with tilted edges.}
	\label{fig:S9}
\end{figure}

The mapping of the trivial configuration to a spin-0 chain and of the topological configuration to a spin-1 chain (see Fig.~\ref{fig:fig1} of the main text) is investigated numerically in ED calculations on a system of length $L=5$ at zero temperature with $M^z = 0$ (see Fig.~\ref{fig:S9}). As expected, the rung singlet fraction increases monotonically with $\Jp/\Jpl$ and approaches 1 for $\Jp\gg\Jpl$. Remarkably the triplet fraction along the diagonals is always high ($\geq80\%$) and reaches its maximum at $\Jp/\Jpl \sim 1$ consistent with \cite{white:1996}. The order parameters in the main text prove that, even when tuning $\Jp/\Jpl$ away from the limiting cases, the system stays within the respective trivial (topological) phase.

\subsection*{Normalisation effects}
The string-only correlator $\gI$ is naturally normalised as it returns 1 for any state with an even number of $|S^z|=1$ in the string. The spin-string correlator $\gS$ only equals 1 in a classical \mbox{spin-1} N{\'e}el state. Any spin-1 state with rotational symmetry has $\gS<1$ due to the presence of some $S^z=0$ at the end of the string. A meaningful normalisation is given by $\gtS = \eta ~ \gS$ with $\eta^{-1}=\left\langle \big|\hat{S}^z_k \big| \big|\hat{S}^z_{k+d}\big|\right\rangle$ describing the probability that neither endpoint of the string has spin $S^z_k=0$. In direct analogy to Bayesian conditional probabilities, $\gtS$ describes the string-correlation between $|S^z|=1$ spins and thus $|\gtS|=1$ for the AKLT state.

\begin{figure}[t]
	\centering
	\includegraphics{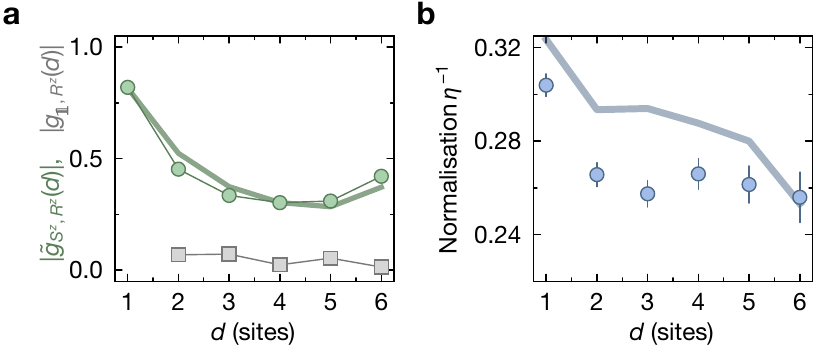}
	\caption{\textbf{Normalisation of string correlators.} \textbf{a}, Normalised correlator $\gtS$ compared to $\gI$ in the topological regime measured at length $L=7, \Jp/\Jpl = 1.3(2)$ with $M^z = 0$ and \textbf{b}, corresponding values for $\eta^{-1}$. Shaded lines are ED calculations at $S/N = 0.45\,\kB$, $\Jp/\Jpl = 1.2$, $L=7$, $M^z = 0$. At small distances $d$ the normalisation is higher due to correlation between the unit-cells.
	}
	\label{fig:S11}
\end{figure}

The application of this normalisation to the data of Fig.~\ref{fig:fig2} is shown in Fig.~\ref{fig:S11}a with the explicit values for $\eta$ given in Fig.~\ref{fig:S11}b. For $d\gg1$, it is equivalent to the normalisation used in \cite{white:1996}. 
We use the same normalisation in the Hubbard regime (see Fig.~\ref{fig:fig4} of the main text).

\subsection*{Finite system size and temperature}
\begin{figure}[b]
	\centering
	\includegraphics{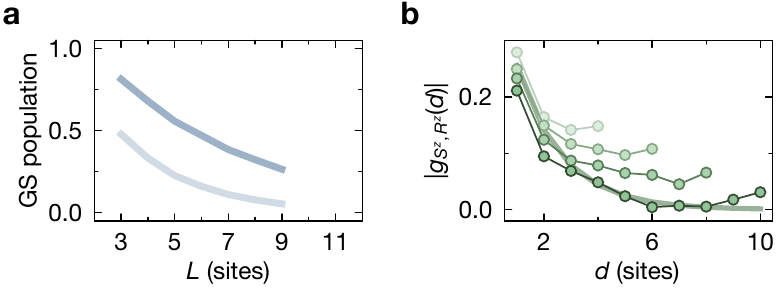}
	\caption{\textbf{Finite temperature and finite length effects.} \textbf{a}, Numerical ground state (GS) population for a fixed entropy of $S/N = 0.3\,\kB$ ($0.45\,\kB$) in (light) blue. The ground state population decreases with length for all finite temperatures. \textbf{b}, Experimentally measured string correlator $\gS(d)$ for system sizes $L= 5,7,9$ and $11$ with $J_\perp/J_\parallel = 1.3(2)$. The shaded line shows finite temperature, infinite length DMRG calculations at $\Jp/\Jpl = 1.3,\, T = 0.9\Jpl$.}
	\label{fig:S7}
\end{figure}

In the thermodynamic limit, the SPT phase only exists at strictly zero temperature (i.e. the spin-string correlator $\gS(d\to \infty)\neq 0$). The reason is that there are only four ground states, but infinitely many excited states just above the energy gap resulting in infinitely many singlets that can be broken. In our experimental setup, we can nevertheless observe the characteristics of this phase. The finite length of our system limits the number of low energy states available, such that even at a temperature around the gap energy, the ground state is still largely populated (see Fig.~\ref{fig:S7}). The colder the system, the longer the length at which the ground state still dominates. Fig.~\ref{fig:S7}a shows the ground state population for short system lengths at $J_\perp/J_\parallel=1.2$ for $S/N = 0.3\,\kB$ and $S/N = 0.45 \,\kB$ corresponding to a temperature of $T=0.6\,\Jpl,~1.2\,\Jpl$. 
The ground state population quickly drops as the number of available states increases. The effect of the reduced ground state occupation can be seen in the measured string correlator $\gS(d)$ in systems of different length (see Fig. \ref{fig:S7}b). The system of $L= 11$ shows a much lower value for the string correlator even for short distances, where smaller systems show significantly higher correlations. This restricts the system size up to which signatures of the Haldane phase can be detected in an experiment but even below this system size, all qualitative features of the zero-temperature phase are already present. Furthermore, to show that the signal is not dominated by the small system size, we use DMRG calculations at infinite length at $T=0.9\Jpl\,\Jp/\Jpl = 1.3$ (shaded line in Fig.~\ref{fig:S7}b). 

\subsection*{Finite size offset}
\begin{figure}[t]
	\centering
	\includegraphics{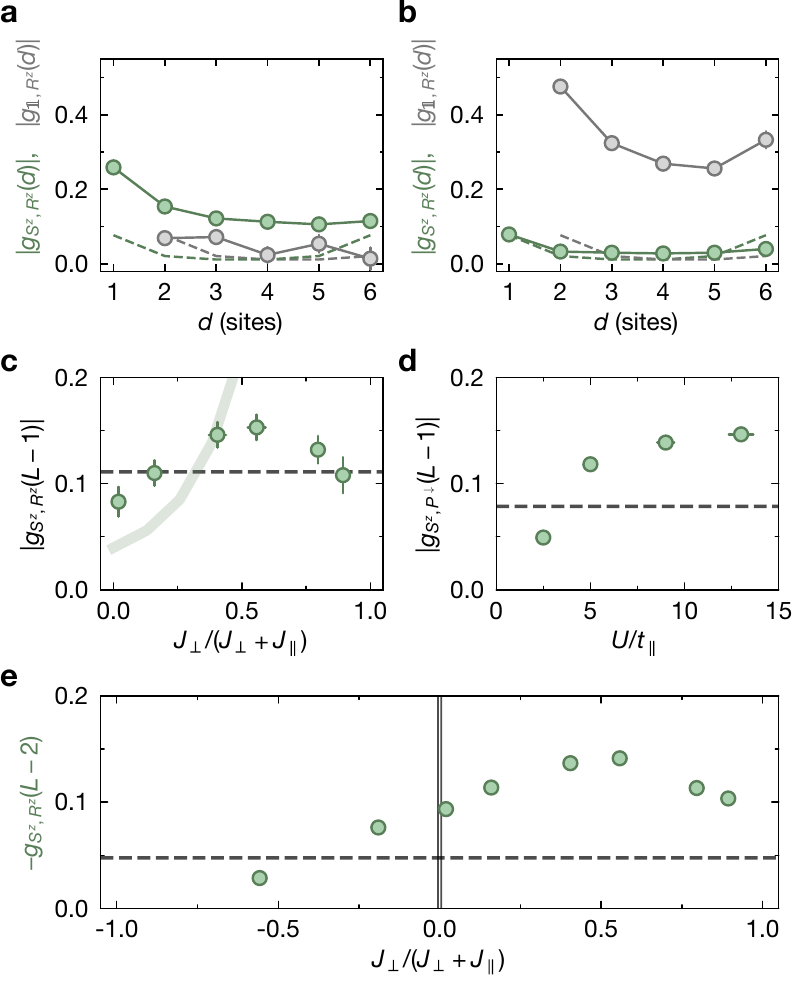}
	\caption{\textbf{Finite size offset of the string correlator.} Finite size offset calculated as a function of string distance for the topological (\textbf{a}) and trivial (\textbf{b}) regime. Dashed lines are numerical calculations, solid lines are measured data presented also in Figure \ref{fig:fig2}. All values are for $L=7$, $M^z = 0$ and $\Jp/\Jpl = 1.3(2)$. 
	\textbf{c}, The finite size offset of the spin string correlator for the infinite temperature Heisenberg model (dashed line) of a topological system with $L=5,~M^z = 0$ for different $\Jp/\Jpl$. The shaded line shows the zero temperature value from ED calculations. \textbf{d}, Similarly, the finite size offset for the Hubbard model compared to the novel spin string correlator (cf. Fig.~\ref{fig:fig4}). 
	\textbf{e}, String correlator $g_{S^z,R^z}$ at distance $d=L-2$. For all coupling parameters $J_\perp/J_\parallel$ in the topological regime, the measured value clearly exceeds the infinite temperature finite size offset (dashed line).
	}
	\label{fig:S8}
\end{figure}
In a finite-size system, the string correlator $\gS$ does not approach zero even when the temperature increases to infinity if the magnetisation is fixed. 
The lack of free fluctuations of $M^z$ introduces correlations in the system even for a random distribution of the spins\cite{hilker:2017}. 
These correlations, which can be derived from combinatorial considerations, do not depend on the edge termination or coupling parameters. Fig.~\ref{fig:S8} investigates the effect of this offset on our measurements. As can be seen in Fig.~\ref{fig:S8}a,b, the offset mostly takes sizeable values at the shortest and longest distances of the system. But even at these points, our signal clearly exceeds the offset. Fig.~\ref{fig:S8}c shows the string correlator as a function of the coupling strength (as in Fig.~\ref{fig:fig3}a). Interestingly the string correlation value can coincide with or even be lower than the infinite temperature offset for couplings far away from the symmetric point of $\Jp \approx \Jpl$. This is however not an artefact of the measurement but is also reflected in the zero temperature ED calculations (shaded line). Thus it is not meaningful to simply subtract this offset. Similarly we show the offset for the Hubbard regime in Fig.~\ref{fig:S8}d. \\
As noted before, the offset is considerably smaller for intermediate string lengths. Therefore we investigate the value of the string correlator for $d=L-2$ (see Fig.~\ref{fig:S8}e). In this case, the offset is considerably smaller than the measured values for any coupling on the topological side. Therefore our system is not dominated by its finite system size but exhibits the topological properties of the bulk characteristic to the Haldane phase.

\begin{figure*}[t]
	\centering
	\includegraphics{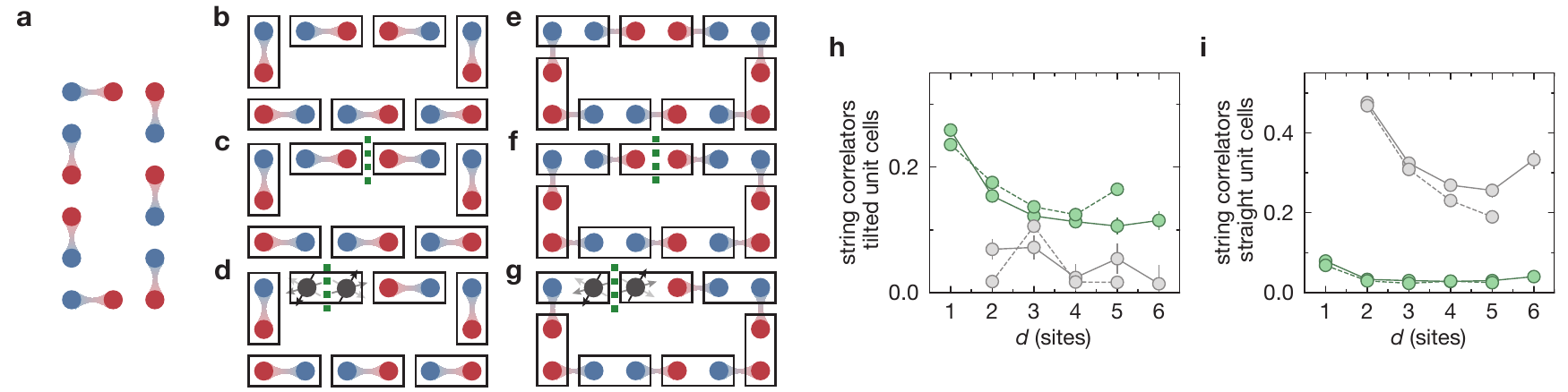}
	\caption{\textbf{Illustration of the relationship between edges and unit cells.} \textbf{a}, Ring of singlets without any predefined unit cell. No clear statement about the topological or trivial phase can be made here. On the right, the first row uses unit cells aligned with the singlet bonds (which corresponds to vertical unit cells in the experiment). The second row has unit cells connecting different singlets, i.e. diagonal unit cells. Both start with periodic boundary conditions (\textbf{b}, \textbf{e}) where no clear distinction between the topological and the trivial regime can be made. Panel \textbf{c} and \textbf{f} cut the system into a chain while leaving the singlets intact. This corresponds to the system with straight edges. Meanwhile \textbf{d} and \textbf{g} cut singlets as in the system with tilted edges. Note that only in \textbf{c} and \textbf{g} the definition of the unit cell agrees with the cut of the system leading to the terminology of a trivial chain for straight edges and topological chain for tilted edges. \textbf{h} (\textbf{i}) show the string correlators $g_{S^z, R^z}$ (green) and $g_{\mathds{1},R^z}$ (grey) for diagonal (vertical) unit cells. Solid lines correspond to edges matching the unit cell (i.e. tilted edges with for the diagonal unit cell and vice versa), dashed lines to opposite orientation such that there are half unit cells at the edges.  The choice of edge termination only has a minor effect on the signal. }
	\label{fig:S14}
\end{figure*}

\subsection*{Unit cell and edge effects} 
In the main text we present two different edge terminations as realisations of the topological and trivial phase. The influence of the edges does however not fully determine the behaviour of the two string correlators. This is to be expected as the string correlators probe the bulk of the system which one could also calculate in an infinite system. Only in conjunction with a chosen unit cell does it make sense to  characterise the system via a specific phase. \\
As an illustrative example we use a singlet chain (see Fig.\,\ref{fig:S14}), where every spin-1/2 particle is paired into a singlet with a fixed neighbour. In the case of periodic boundary conditions (a,\,b,\,e), the topology of the system with a two-site unit cell is only set by the choice of unit cell: Unit cells around the the singlets result in a trivial bulk, while the shifted unit cell leads to a topological Haldane phase. We stress that on the level of spin-1/2 particles both states are identical and only the different choices of pairing in the analysis lead to the different topologies. \\
By cutting, the ring can be turned into a chain with edges. But because there is no long-range entanglement in the system, the bulk properties stay unaffected by the cut i.e. the topology is still set by the choice of unit cell and not by the position of the cut. Only one of the two choices of units cell, however, agrees with the cut such that no unit cell is split (c,\,g). By providing such a natural choice of unit cell, the edge is linked to the topology of the bulk. When only considering the bulk it is also possible to choose the opposite unit cell, which disagrees with the cut, leading to the opposite string-correlation results (d,\,f).\\ 

In the main text, we presented results for the natural choice of units cells (vertical for straight edges, diagonal for tilted edges). In Figure.~\ref{fig:S14}h,i, we compare the string correlators \gS and \gI with both edge terminations at fixed unit cells. In tilted unit cells (h), we always find a topological bulk, while straight unit cells (i) show the trivial correlations. The physical presence or absence of an unpaired spin-1/2 does not affect the bulk significantly. This is consistent with the discussion in the last paragraph and demonstrates, that, even in our relatively small systems, the properties of the bulk are independent of the edge.\\

\subsection*{Microscopy of edge states}
Due to our microscopic resolution, we can study the magnetisation pattern within the localised edge state in detail. 
In Fig.~\ref{fig:S4}, the spatial magnetisation distribution of a system with strong rung coupling is compared to the balanced situation ($\Jp \approx \Jpl$). As the $M^z =1$ sector has positively polarised edge states at $T=0$, the magnetisation maps directly reveal the structure of the state: the excess magnetisation dominantly sits at the edge but leaks into the bulk where it induces a staggered magnetisation pattern close to the edge due to the AFM spin coupling. How far the correlation extends into the bulk is set by the leg coupling $\Jpl$ relative to the bulk gap. At finite temperature, there is, in addition, some homogeneous magnetisation of the bulk. In Fig.~\ref{fig:fig3} of the main text, we show the unit-cell average of this data.

\subsection*{Edge state splitting}
For finite system length, the four ground states of the SPT phase are not truly degenerate but exhibit a finite energy splitting $\delta$. This energy splitting arises from an exponentially suppressed but non-zero overlap between the edge states, coupling them into singlet and triplet states. Whether the singlet state or the triplet state is lower in energy depends on the parity of the system length $L$, whereas the energy splitting depends on the system length directly (see Fig.~\ref{fig:S5}a). 

\begin{figure}[t]
	\centering
	\includegraphics{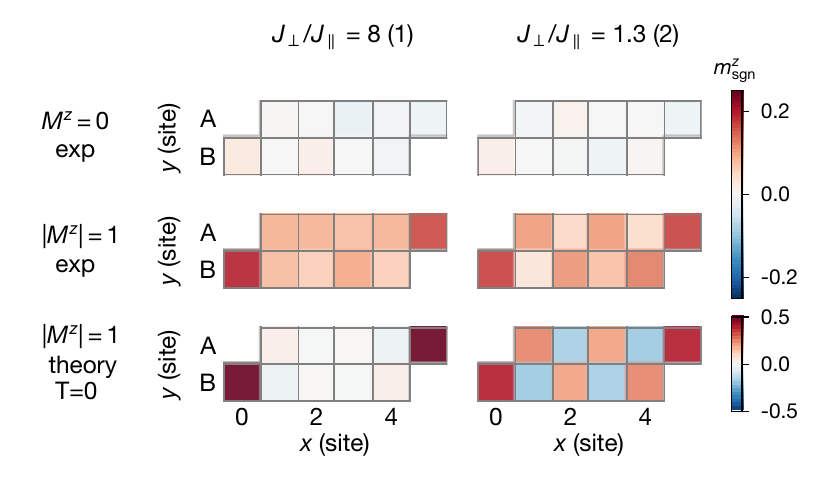}
	\caption{\textbf{Spatial magnetisation distribution in the spin-1/2 ladder.} Experimental magnetisation maps $m^z_{x,y}$ for different $J_\perp/J_\parallel$ are shown for $M^z=0$ in the first row, $ |M^z|=1$ in the second row and zero temperature Heisenberg model at $ |M^z|=1$ in the last row. The plotted quantity is $m^z_\text{sgn} = m_{x,y}^z \cdot M^z$, flipping the sign in the $M^z=-1$ sector.}
	\label{fig:S4}
\end{figure}

\begin{figure}[t]
	\centering
	\includegraphics{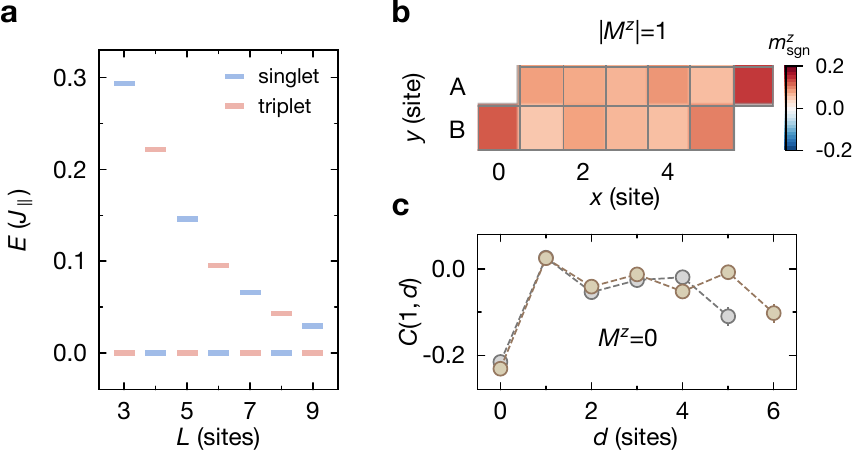}
	\caption{\textbf{Edge states for even and odd system length.} \textbf{a}, Ground-state energies of the Haldane phase for different system lengths. The lowest-lying state alternates between the singlet and triplet state, depending on the parity of the system length. The data is produced via exact diagonalisation of the spin-1/2-Heisenberg ladder. \textbf{b}, The magnetisation map of our system at total magnetisation $|M^z|=1$ at $L= 6$ and $\Jp/\Jpl = 1.3(2)$. It shows the onset of an alternating magnetisation pattern close to the edge sites, but gets lost in the centre of the ladder since the patterns of the two edges have opposite phase. The plotted value $m^z_\text{sgn}$ is the same as in Fig.~\ref{fig:S4}. \textbf{c}, Spin correlations $C(1,d)$ for a system at total magnetisation $M^z = 0$ of $L= 6$ (brown) and of $L= 5$ (grey). Both show a strong nearest-neighbour correlation as well as a strong edge to edge correlation, indicating the edge states of opposite spin for this magnetisation. However, for $L= 6$ the correlations alternate around a finite-size finite-temperature offset, whereas for $L= 5$ they do not alternate because an alternation does not match the length and the negative endpoints.}
	\label{fig:S5}
\end{figure}

Experimentally we cannot observe the splitting directly due to our finite temperature but we explore the underlying spin correlations responsible for the energy splitting:
we investigate the effect of even and odd system length by comparing a system of $L= 6$ (see Fig.~\ref{fig:S5}b,\,c) to one with $L= 5$ (see Fig.~\ref{fig:S4}). 
The staggered magnetisation pattern seen for $L= 5$ is not visible for $L=6$ because the induced spin pattern of both edges are incommensurate with each other leading to the higher energy of the triplet state. 
The opposite is true in the $M^z = 0$ sector, where we find stronger alternating patterns in the even ladder length compared to the odd one (see Fig.~\ref{fig:S5}c around $d=3,4$). Here we analyse the spin-correlations $C(1,d) = 4\langle \hat{S}^z_{i,j}\hat{S}^z_{i+1,j+d}\rangle$ because for $M^z=0$ the local magnetisation is zero everywhere. 
These observations illustrate how the alternation between singlet and triplet ground states with $L$ is linked to the AFM polarisation due to the edges.

\enlargethispage\baselineskip

\subsection*{Localisation length}
We next relate the experimentally extracted edge state decay length $\xi$ to the localisation of the edge at zero temperature. 
Numerically, the length over which the edge modes delocalise can be readily extracted from the aforementioned energy splitting $\delta$
\begin{equation*}
| \delta | \sim e^{-L/\xi}.
\end{equation*}
The experimental edge decay length is determined, however, using the staggered magnetisation that arises for $|M^z| =1$
\begin{equation*}
| m^z(k) | \sim e^{-k/\xi},
\end{equation*}
where $k$ denotes the position of the unit cell along the chain. 

\begin{figure}[t]
	\centering
	\includegraphics{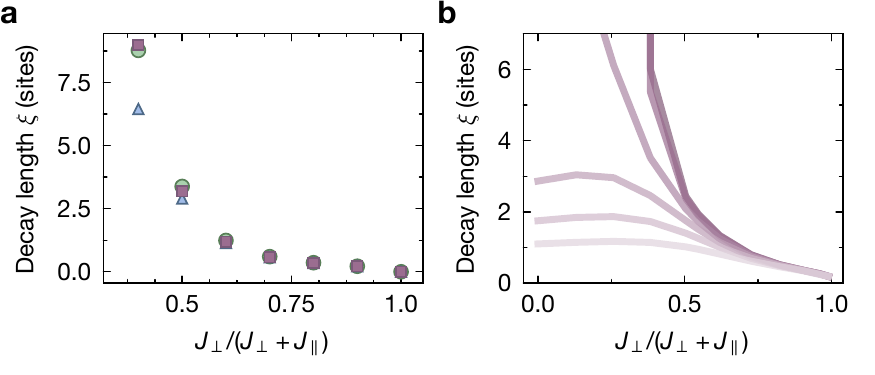}
	\caption{\textbf{Decay length comparison.} \textbf{a}, The decay length of several quantities is determined using DMRG for a system of $L>50$, and $U/t_\parallel = 13$. The purple squares show the decay length of the magnetisation pattern, as it is also evaluated on the experimental data. The green circles show the edge localisation length derived from the edge state splitting and the blue triangles show the bulk correlation length. When markers are not visible they coincide with the purple squares. \textbf{b}, Finite temperature results for a system with $L= 5$ and $S/N = (0-0.5)\,\kB$ (from dark to light purple). The decay length is calculated from the magnetisation pattern in the Heisenberg model using ED.}
	\label{fig:S6}
\end{figure}

These two approaches are numerically compared in Fig.~\ref{fig:S6}a using DMRG in a system with $L=100$, $U/t_\parallel=13$ at zero temperature. The decay length is evaluated from the energy splitting in the different spin sectors $M^z\in\{0,1\}$. There, we choose $L\in[4,8,...,52]$ and a bond dimension $\chi=1000$ keeping the maximal energy truncation error below $10^{-7}$. For the edge magnetisation, we calculated the ground state in the sector $M^{z}=1$ for $L=100$ and a bond dimension $\chi=1000$ again with an error below $10^{-7}$ for all parameters. Both quantities agree with deviations of less than 6\% of their values, confirming the validity of our method to extract the localisation length from the experimental data (see Fig.~\ref{fig:fig3}). 
We furthermore compute the bulk correlation length, which is a direct measure of the bulk gap. It can easily be obtained from the ground state of the infinite system~\cite{schollwock:2011} and coincides with the former length scales for most parameters. 

However, the decay length obtained from the staggered magnetisation shows a strong temperature dependence (see Fig.~\ref{fig:S6}b), whereas the edge state splitting is a property of the spectrum, and thus independent of temperature. 
The staggered magnetisation arises from the antiferromagnetic correlation of the spin, which decreases with temperature. We find that the edge state cannot delocalise beyond the thermal coherence length of the system. 
The finite temperature decay length thus follows the zero temperature decay length when it is small (large $\Jp/\Jpl$) but saturates in the low $\Jp/\Jpl$ regime at an upper bound given by the temperature of the system.

\subsection*{Symmetry fractionalisation}
The key to understanding one-dimensional SPT phases is the notion of symmetry fractionalisation \cite{pollmann:2010,chen:2011c}. If $\hat{U} = \prod_n \hat{U}_n$ is an on-site unitary symmetry, it can be argued that if the ground state $|\psi\rangle$ is symmetric under $\hat{U}$ and has a finite correlation length, then 
\begin{equation}
\prod_{k=m}^n \hat{U}_k |\psi\rangle = \hat{U}_L \hat{U}_R |\psi\rangle, \label{eq:symfrac}
\end{equation}
where $\hat{U}_L$ ($\hat{U}_R$) is a unitary operator which is exponentially localised to the left (right) of the block of sites $k=m,m+1,\cdots,n-1,n$. Moreover, if $\hat{V}$ is another symmetry, then the group relations between the fractionalised symmetries $\hat{V}_L$ and $\hat{U}_L$ are the same as those between the bulk symmetries $\hat{V}$ and $\hat{U}$ up to potential phase factors. As an example, suppose that $\hat{U}$ and $\hat{V}$ commute and suppose that $\hat{U}_{L,R}$ are bosonic operators (which is the case for all symmetries considered in this work), then
\begin{align}
\mathds{1} &= (\hat{U}_L\hat{U}_R)(\hat{V}_L \hat{V}_R)(\hat{U}_L\hat{U}_R)^{-1}(\hat{V}_L \hat{V}_R)^{-1} \\
&= \left( \hat{U}_L \hat{V}_L \hat{U}_L^{-1} \hat{V}_L^{-1} \right) \left( \hat{U}_R \hat{V}_R \hat{U}_R^{-1} \hat{V}_R^{-1} \right).
\end{align}
Since the fractionalised operators on the left and right have disjoint support (up to exponentially small overlaps), we have that $\hat{U}_L \hat{V}_L \hat{U}_L^{-1} \hat{V}_L^{-1} = e^{i\alpha} \mathds{1}$, i.e., $\hat{U}_L \hat{V}_L = e^{i\alpha} V_L \hat{U}_L$: the fractionalised symmetries commute up to a phase. More generally, the group relations of the fractionalised symmetries define a \emph{projective representation} of the symmetry group. Some of these phase factors can be gauged away by redefining $\hat{U}_L,\hat{U}_R \to e^{i\beta} \hat{U}_L,e^{-i\beta} \hat{U}_R$, while other phase factors are invariant. The collection of such invariant phase factors define the so-called second group cohomology class $H^2(G,U(1))$: any non-zero element in this class represents a non-trivial SPT phase. The Haldane SPT case corresponds to where the bulk symmetry group is $SO(3)$ (or its $\mathbb Z_2\times \mathbb Z_2$ subgroup) and the fractionalised symmetries form $SU(2)$ (or its quaternion subgroup).

\subsection*{From symmetry fractionalisation to edge modes}
Note that a non-trivial projective representation for the fractionalised symmetries automatically implies edge modes: for open boundaries, one can consider Eq.~\eqref{eq:symfrac} as acting on the whole system, which moreover implies that $\hat{U}_L$ and $\hat{U}_R$ are genuine symmetries of the ground state. Since a projective representation can never act on a one-dimensional Hilbert space, there must be a degenerate zero-energy Hilbert space associated to the boundaries (in other words, the ground state cannot be a simultaneous eigenstate of all fractionalised symmetries).

\subsection*{From symmetry fractionalisation to string order parameters \label{subsec:string}}
For an on-site symmetry $\hat{U} = \prod_n \hat{U}_n$, consider a string operator $\hat{\mathcal O}^\dagger_m \hat{U}_{m+1} \cdots \hat{U}_{n-1} \hat{\mathcal O}_n$ with endpoint operator $\hat{\mathcal O}_n$ and $n-m$ being much larger than the correlation length. Using symmetry fractionalisation, its expectation value can be expressed as
\begin{equation}
\langle \hat{\mathcal O}^\dagger_m \hat{U}_{m+1} \cdots \hat{U}_{n-1} \hat{\mathcal O}_n \rangle = \langle \hat{\mathcal O}^\dagger \hat{U}_L \rangle \langle \hat{U}_R \hat{\mathcal O} \rangle.
\end{equation}
Hence, long-range order (LRO) for this string operator is equivalent to $\langle \hat{\mathcal O}^\dagger \hat{U}_L \rangle \neq 0$. In the assumption that the ground state is symmetric (i.e., no spontaneous symmetry breaking), a local operator can only have a non-zero expectation value if it is neutral under the symmetry group. This means that LRO is only possible if $\hat{\mathcal O}$ is chosen to have the same symmetry charges as the fractionalised symmetry $\hat{U}_L$ (i.e., finding the right $\hat{\mathcal O}$ allows to infer the projective representation of the fractionalised symmetries). This is how string order parameters can be used to diagnose an SPT phase \cite{pollmann:2012a}. 

As an example, consider an on-site $\mathbb Z_2\times \mathbb Z_2$ symmetry, such as the $\pi$-rotations $\hat{R}_x$ and $\hat{R}_z$ for an integer spin chain. The fractionalised symmetries either commute or anticommute: $\hat{R}_x^L \hat{R}_z^L = \pm \hat{R}_z^L \hat{R}_x^L$; the anticommuting case gives rise to the topological Haldane phase. In particular, this means that $\hat{R}_x \hat{R}_z^L \hat{R}_x^\dagger = \hat{R}_x^L \hat{R}_z^L \left( \hat{R}_x^L \right)^\dagger = - \hat{R}_z^L$. Hence, a string for $\hat{R}_z$ can only have LRO if the endpoint operator is odd under $\hat{R}_x$. This explains the well-known string order parameter for the Haldane SPT phase: $\cdots \hat{R}_z \hat{R}_z \hat{R}_z \hat{S}^z$. Conversely, the trivial phase can only have LRO for the $\hat{R}_z$ string if the endpoint is even under $\hat{R}_x$; choosing the endpoint operator to be the identity operator does the trick.

If one tunes away from the Mott limit, the distinction between these two cases breaks down. The integer spin chain now does not have $SO(3)$ symmetry but rather $SU(2)$ symmetry. More concretely, $\hat{R}_x$ and $\hat{R}_z$ are no longer $\mathbb Z_2$ symmetries: they square to fermion parity symmetry $\hat{P}$, not to the identity. (In the Mott limit, fermion parity is a classical number in each unit cell, giving rise to an effective spin chain.) This means that the fractionalised symmetries now obey $\hat{R}_x^L \hat{R}_z^L = \hat{P}^L \hat{R}_z^L \hat{R}_x^L$, which allows to adiabatically connect the spin chains where $\hat{P}^L = \pm \mathds{1}$ as has been demonstrated before \cite{anfuso:2007,moudgalya:2015,verresen:2021}.

\subsection*{A novel string order parameter for anti-unitary symmetry \label{subsec:novel}}
Let $\hat{T}$ be an anti-unitary symmetry. Analogous to Eq.~\eqref{eq:symfrac}, its symmetry fractionalisation can be written as $\hat{T} = \hat{U}_L \hat{U}_R \tilde K$, where $\tilde K$ is complex conjugation defined with respect to a basis that factorises between left and right (see \cite{pollmann:2010,verresen:2017} for details). A single anti-unitary $\mathbb Z_2^T$ symmetry can protect a non-trivial SPT phase. More precisely, $\hat{T}^2$ implies that $\hat{U}_L \tilde K \hat{U}_L \tilde K = \hat{U}_R \tilde K \hat{U}_R \tilde K = e^{i\theta} = \pm 1$; the case $\theta = \pi$ is the topological Haldane phase, where the edge mode is a Kramers pair under $\hat{T}$.

Usually, it is said that there is no simple string order parameter to detect such an SPT phase protected by $\mathbb Z_2^T$. The simplest ``string order'' is rather involved, requiring two copies of the system and a partial swap \cite{pollmann:2012a}. Here, we show that a conventional string order parameter can be constructed if the system has an \emph{additional} unitary $\mathbb Z_2$ symmetry, which we denote as $\hat{P}$.

Suppose that we consider phases which are not protected by the combined symmetry $\hat{P}\hat{T}$. We now show that the string order parameter for a $\hat{P}$-string, i.e., $\hat{\mathcal O}^\dagger_m \hat{P}_{m+1} \cdots \hat{P}_{n-1} \hat{\mathcal O}_n$, can be used to diagnose whether the phase is in a trivial or topological phase with respect to the anti-unitary symmetry $\hat{T}$. To see this, first note that the phase being trivial with respect to $\hat{P}\hat{T}$ implies that if $\hat{P} = \hat{P}_L\hat{P}_R$ is the fractionalisation of $\hat{P}$, then $\hat{P}\hat{T} = \hat{P}_L \hat{U}_L \hat{P}_R \hat{U}_R \tilde K$ must obey $\hat{P}_L \hat{U}_L \tilde K \hat{P}_L \hat{U}_L \tilde K = +1$. Moreover, since $\hat{P}^2=1$, we can choose $\hat{P}_L^2 = 1$, such that
\begin{equation}
\hat{P}_L = \hat{U}_L \tilde K \hat{P}_L \hat{U}_L \tilde K.
\end{equation}
We thus obtain $\hat{T} \hat{P}_L \hat{T} = \hat{U}_L \hat{U}_R \tilde K \hat{P}_L \hat{U}_L \hat{U}_R \tilde K = \left( \hat{U}_L \tilde K \hat{P}_L \hat{U}_L \tilde K \right) \left( \hat{U}_R \tilde K \hat{U}_R \tilde K\right) = e^{i\theta} \hat{P}_L$. Hence, whether $\hat{P}_L$ commutes or anticommutes with $\hat{T}$ encodes what phase we are in. Using the reasoning of the previous section, one can conclude that the string order parameter for $\hat{P}$ with endpoint operator $\hat{\mathcal O}$ (where we choose $\hat{\mathcal O}$ to be hermitian) has long-range order if $\hat{T} \hat{\mathcal O}_n \hat{T} = e^{i\theta} \hat{\mathcal O}_n$.\\

\subsection*{Application to Hubbard chain}
We have already explained why away from the Mott limit, we can no longer rely on the conventional string order parameter to characterise an SPT phase. However, as pointed out in \cite{verresen:2017}, the (bond-alternating) Hubbard chain is still in a non-trivial SPT phase protected by an anti-unitary $\mathbb Z_2^T$ symmetry. This uses the bipartite structure of our model (which is evident in the fact that the system lives on a ladder), leading to the symmetry as defined by
\begin{equation}\label{eq:sym}
		\hat T: c_{x,y,s} \leftrightarrow (-1)^{x+y} c_{x,y,s}^\dagger
\end{equation}
where $x = [0,L]$, $y = 0,1$ (corresponding to A or B) and $s = \uparrow, \downarrow$. Here the bipartite property is encoded in the factor $(-1)^{x+y}$ as the ladder Hamiltonian only couples sites with opposite parity.
This follows directly from combining two facts: (i) it is a $\mathbb Z_2^T$ symmetry of the model, even away from the Mott limit, and (ii) in the Mott limit, it can be argued that it coincides with spinful time-reversal symmetry \cite{verresen:2017}, which is known to protect the Haldane SPT phase \cite{pollmann:2010}.

To construct a string order parameter for this phase, we use the result obtained in the previous section. In particular, consider the additional $\mathbb Z_2$ symmetry $\hat{P}_\downarrow$, which is the fermion parity of the down-spin species. In the Mott limit, this symmetry becomes indistinguishable from $\hat{R}_z$. From this, we learn that $\hat{P}_\downarrow \hat{T}$ does not protect the SPT phase (which was a condition that we assumed in the previous section). As derived above, this implies that the string operator associated to $\hat{P}_\downarrow$ can be used to read off the topological invariant: we are in the Haldane SPT (trivial) phase if the string has long-range order for an endpoint operator that is odd (even) under $\hat{T}$. For instance, for the topological phase, we can thus choose the endpoint operator $\hat{\mathcal O}_n = \hat{S}^z_n = \frac{1}{2} \left( \hat{c}_{n,\uparrow}^\dagger \hat{c}_{n,\uparrow}^{\vphantom \dagger} - \hat{c}_{n,\downarrow}^\dagger \hat{c}_{n,\downarrow}^{\vphantom \dagger} \right)$, which is odd under the above anti-unitary $\hat{T}$ symmetry.

\end{document}